# Circularly Polarized Luminescence Without External Magnetic Fields from Individual CsPbBr$_3$ Perovskite Quantum Dots

Virginia Oddi[1,2,#], Chenglian Zhu[2,3,#], Michael A. Becker[1,$], Yesim Sahin[2,3], Dmitry N. Dirin[2,3], Taehee Kim[2,3], Rainer F. Mahrt[1], Jacky Even[4], Gabriele Rainò[2,3,*], Maksym V. Kovalenko[2,3,*], Thilo Stöferle[1,*]

[1] IBM Research Europe – Zurich, Säumerstrasse 4, 8803 Rüschlikon, Switzerland.

[2] Institute of Inorganic Chemistry, Department of Chemistry and Applied Biosciences, ETH Zurich, 8093 Zurich, Switzerland.

[3] Laboratory for Thin Films and Photovoltaics, Empa, Swiss Federal Laboratories for Materials Science and Technology, 8600 Dübendorf, Switzerland.

[4] Université de Rennes, INSA Rennes, CNRS, Institut FOTON - UMR6082, 35000 Rennes, France.

**\* Corresponding Authors:** rainog@ethz.ch, mvkovalenko@ethz.ch, tof@zurich.ibm.com
[#] These authors equally contributed to this work.
[$] Present address: Zeiss SMT, Oberkochen, Germany

Lead halide perovskite quantum dots (QDs), the latest generation of colloidal QD family, exhibit outstanding optical properties which are now exploited as both classical and quantum light sources. Most of their rather exceptional properties are related to the peculiar exciton fine-structure of band-edge states which can support unique bright triplet excitons. The degeneracy of the bright triplet excitons is lifted with energetic splitting in the order of millielectronvolts, which can be resolved by the photoluminescence (PL) measurements of single QDs at cryogenic temperatures. Each bright exciton fine-structure-state (FSS) exhibits a dominantly linear polarization, in line with several theoretical models based on the sole crystal field, exchange interaction and shape anisotropy. Here, we show that in addition to a high degree of linear polarization, the individual exciton FSS can exhibit a non-negligible degree of circular polarization even without external magnetic fields by investigating the four Stokes parameters of the exciton fine-structure in individual CsPbBr$_3$ QDs through Stokes polarimetric measurements. We observe a degree of circular polarization up to ~38%, which could not be detected by using the conventional polarimetric technique. In addition, we found a consistent transition from left- to right-hand circular polarization within the fine-structure triplet manifold, which was observed in magnetic field dependent experiments. Our optical investigation provides deeper insights into the nature of the exciton fine-structures and thereby drives the yet-incomplete understanding of the unique photophysical properties of this novel class of QDs, potentially opening new scenarios in chiral quantum optics.

Circular dichroism is the phenomenon of differential absorption of right- and left-hand circularly polarized light.[1] The effect is observed in the absorption bands of chiral molecules whose symmetry does not allow them to be superimposed on their mirror image.[2] While circular dichroism probes the ground-state properties of materials, circularly polarized luminescence is defined as the right- or left-hand circularly polarized emission and provides information about the luminescent excited state.[3] Circularly polarized emission and circular dichroism are found in inherently chiral systems like chiral molecules,[4,5] chiral supramolecular assemblies,[6] lanthanide ion complexes,[7] transition metal complexes,[8] chiral biomolecular systems,[9] or chiral quantum dots (QDs)/rods.[10] Moreover, chirality of an emitter can be induced by the attachment of chiral ligands,[11,12] or by a surrounding chiral matrix/solvent,[13] leading to the preferred emission/absorption of left- or right-hand circularly polarized light. Differential light-matter interaction of left- and right-hand circularly polarized light can also be induced by static magnetic fields.[14] While non-magnetic methods are primarily used to investigate structural and stereochemical information, magnetically induced circular dichroism and especially circularly polarized luminescence can give important insight into the electronic structure of the emitters.[15] In low-temperature magneto-optical measurements on single emitters *e.g.*, QDs, the type of polarization (linear or circular) elucidates optoelectronic fine-structure properties of excited states.[16–18] Although chiroptical phenomena are extensively studied for fundamental research, there is also a growing interest in advanced photonic technologies that exploit the differential emission/absorption of right- and left-hand circularly polarized light,[5] such as ellipsometer-based tomography[19] or novel LED[20] and display technology.[21] Moreover, circularly polarized luminescence of quantum emitters is central to the promising field of chiral quantum optics,[22] with potential applications in quantum-information processing[23] and quantum simulation.[24]

**Material system**

Cesium lead halide ($CsPbX_3$) perovskite QDs are a rapidly emerging class of colloidal QDs due to their outstanding optoelectronic properties.[25–27] The crystal structure of $CsPbX_3$ perovskite is characterized by a three-dimensional network of corner-sharing $PbX_6$ octahedra (X=Br, Cl, I) with the $Cs^+$ filling the void, exhibiting an orthorhombic structure at low temperature, as shown in left panel of Figure 1a. With fluorescence quantum yields (QY) approaching unity and a wide tunability of the emission wavelength,[28,29] this type of QDs can be utilized in various applications, such as quantum light sources,[30–32] solar cells,[33] lasers,[34] displays,[35] and even scintillators.[36] Specifically at cryogenic temperature, a non-degenerate bright triplet state with three orthogonal dipoles and high oscillator strength has been discovered.[18,26,37,38] In addition, the $CsPbBr_3$ QDs show fast radiative lifetime (~100 ps) with the exciton dephasing time on the order of several tens of picoseconds,[30,39–42] which enables coherent light-matter interaction.[43]

High-resolution transmission electron microscope (HRTEM) image of one $CsPbBr_3$ QDs sample is displayed in the right panel of Figure 1a. These cuboidal QDs possess a side length of $14.0 \pm 1.1$ nm and a photoluminescence (PL) QY of 62% in solution at room temperature. In Figure 1b, we show an exemplary PL spectrum of a single $CsPbBr_3$ QD at cryogenic temperature (4 K), exhibiting doublet exciton fine-structure with an energy splitting of $\Delta = 0.75 \pm 0.08$ meV. As shown in many reports,[18,25,26,38,42,44,45] fine-structure states (FSS) of perovskite QDs possess a high degree of linear polarization, which is typically analyzed by tracking the transmitted PL intensity through a linear polarizer at varying angles. The result of such a measurement is depicted in Figure 1c (for the QD displayed in Figure 1b), where two sublevels exhibit a linear polarization profile along crossed orientation. The solid lines represent the $\sin^2$ fits for the PL intensity trajectory of each FSS peak, from which we could extract the degree of linear polarization, $DOLP = (I_{max} - I_{min})/(I_{max} + I_{min})$. For this QD, we obtained a DOLP of $82.9 \pm 2.2\%$ and $78.6 \pm 4.8\%$ for the low- and high-energy emission peaks (FSS1, FSS2), respectively. It should be noted that, in many cases, the above formula is mistakenly used as a measure of the absolute degree of polarization. However, this is only true if the light does not possess any circularly polarized component. To distinguish between unpolarized and circularly polarized light, more sophisticated polarimetric techniques e.g., Stokes polarimetric measurements, are required, from which the four Stokes parameters provide a complete description of different polarization states of light.

**Measurement technique and setup**

In general, the polarization state of electromagnetic field can be fully described by four measurable quantities known as Stokes parameters *I, M, C,* and *S*.[46,47] To analyze the four Stokes parameters from the PL of our QDs, we used a combination of quarter-wave plate (λ/4) and linear polarizer in the detection path as shown in Figure 2a. As depicted in the inset of Figure 2a, $\phi$ is the angle between the vertical (y-axis) and the fast axis of the quarter-wave plate (λ/4), and α is the angle between the vertical (y-axis) and the transmission axis of the polarizer. By measuring the transmitted light intensity while rotating a polarization optical element, one can determine the degree of linear and circular polarization, orientation, and handedness of the light field.[48] To measure the Stokes parameters with the above-described setup, there are two equivalent techniques: measuring the transmitted intensity while rotating either the polarizer,[49] or the quarter-wave plate.[50] We used the latter one because it is advantageous for not causing additional artifacts with the intrinsic polarization dependence of a grating-based spectrometer. Then the detected intensity as a function of the quarter-wave plate angle, $\phi$, is:[48]

$$I(\phi) = \frac{1}{2}\left[I + \left(\frac{1}{2}M \cdot \cos(2\alpha) + \frac{1}{2}C \cdot \sin(2\alpha)\right) \cdot (1 + \cos(\xi))\right] + \frac{1}{2}[S \cdot \sin(\xi) \sin(2\alpha - 2\phi)] + \frac{1}{4}[[M \cdot \cos(2\alpha) - C \cdot \sin(2\alpha)]\cos(4\phi) + [M \cdot \sin(2\alpha) + C \cdot \cos(2\alpha)]\sin(4\phi)] \cdot (1 - \cos(\xi)) \quad (1)$$

Here, $\xi$ is the retardation phase fixed as $\xi = \frac{\pi}{2}$ (for the λ/4-waveplate), $\alpha$ is the angle of the polarizer, and $I$ is the total intensity of the light beam. Correspondingly, the ratios $\frac{M}{I}$ and $\frac{C}{I}$ represent the degree of linear polarization in horizontal, vertical direction and diagonal $(+45°, -45°)$ direction, respectively. Similarly, $\frac{S}{I}$ represents the degree of circular polarization. In specific, the analyzed light is right-hand circularly polarized (RHCP) if $\frac{S}{I} > 0$, and left-hand circularly polarized (LHCP) if $\frac{S}{I} < 0$. Based on these Stokes parameters, the degree of linear polarization is defined as $DOLP = \frac{\sqrt{M^2 + C^2}}{I}$ and the total degree of polarization is defined as $DOP = \frac{\sqrt{M^2 + C^2 + S^2}}{I}$.

Based on Equation (1), we could predict the intensity modulation for differently polarized light as a function of the λ/4-waveplate angle, $\phi$. In Figure 2b, the calculated intensity modulation for fully linearly polarized light along the horizontal and vertical direction (upper panel) and fully left- ($\sigma^+$) and right-hand ($\sigma^-$) circularly polarized light (lower panel) are shown. Figure 2c shows the same for fully linearly polarized light along the +45°/-45° direction (upper panel) and elliptically polarized light (lower panel). For all cases, the degree of polarization was assumed to be 100% and exclusive (e.g., if DOLP = 100%, DOCP = 0%). The curves for purely-circularly polarized light are π-periodic, whereas purely linearly polarized light exhibits a periodicity of π/2, which enables a clear distinction between the linear and circular polarization components. From the phase and intensity of the linear component, it is possible to extract the orientation and degree of linear polarization. Elliptically polarized light includes the components of both linear and circular polarization, and consequently, shows a combination of π/2- and π-periodicity. In particular, the elliptically polarized light shown exemplarily in Figure 2c is characterized by a degree of linear polarization of $DOLP = \frac{\sqrt{M^2+C^2}}{I} = 84.0\%$, and a degree of circular polarization of $DOCP = \frac{|S|}{I} = 54.2\%$, with a total degree of polarization of $DOP = 100\%$. Since $\frac{S}{I} < 0$, the circularly polarized component is left-handed.

A critical aspect to address before performing the polarimetric measurements is to test if the setup introduces any arbitrary modulation, *e.g.*, unwanted birefringent phase shifts occurring often when the light passes through optical coatings. In this regard, each optical element in the detection path should be tested in order to safely exclude experimental artifacts.[51,52] For the setup calibration, we guided a laser beam at 532 nm (near the QD emission wavelength) with a well-defined polarization through the cryostat to mimic the optical path of the QD's PL: through a microscope objective followed by all the optical elements in the detection path, including beam-splitter, long-pass filter, mirrors and lenses. Consequently, the transmitted laser light was analyzed through a rotating λ/4-waveplate and a fixed polarizer as a function of the λ/4-waveplate angle, $\phi$. For more details of the setup, please see the Supporting Information. First, with a combination of linear polarizer and λ/4-waveplate in the excitation path, we confirmed a well-defined linear and circular polarization prepared from our laser output (Figure 2d and S1a for the setup scheme). Next, we analyzed this well-defined light by guiding it through the optical components (PL detection path), as mentioned above (Figure 2e and S1b for the setup scheme). By fitting the intensity traces with Equation (1), we could quantify $\frac{S}{I}$ (DOCP) and DOLP from the Stokes parameters and compare these values before and after going through the detection path of our experimental setup. The prepared laser light originally exhibiting either near-unity DOLP (100%) or DOCP (99.7%) maintained the DOLP and shows only a very marginal change (1.4%) on DOCP after passing through the detection path, respectively. Thus, we validated that our setup introduces negligible modification to the light field passing through, and that the polarization properties can be analyzed free from significant experimental artifacts (at least within the error range of ±1.4%).

Hereafter, we investigated the polarization properties of the exciton FSSs of individual CsPbBr$_3$ QDs at 4 K, utilizing the well-calibrated setup for Stokes polarimetric measurements. We found that these perovskite QDs, capped with non-chiral zwitterionic ligands, exhibit a substantial degree of circularly polarized emission that may reach up to ~38%, which is technically not possible to detect via the standard polarimetric method. Our optical investigation provides deeper insights into different polarization components of the exciton fine-structure that have been poorly studied up to now, which may potentially open new research in chiral quantum optics based on inexpensive, solution-processable, and wavelength-tunable all-inorganic perovskite QDs.

**Measurement results**

In Figure 3a and c, we show two examples of single CsPbBr$_3$ QDs exhibiting clear FSSs with two and three emission peaks, respectively, measured without polarizer or λ/4-waveplate (see corresponding PL time-series in Figure S2a and S2c). During the PL time-series measurements, on timescales of tens of seconds, we do not observe significant changes of the emission intensity due to fluorescence blinking,[25,53–55] which ensure the solidity of Stokes polarimetric measurements. During the Stokes polarimetric measurements, to resolve all the emission peaks and average out intensity fluctuations, we set an integration time of 5 s and 10 s for the QD in Figure 3a and c, respectively. From low to high emission energies, we denote the exciton fine-structure states as FSS1, FSS2 for doublet and FSS1, FSS2, FSS3 for triplet. The QD exhibiting two emission peaks in Figure 3a showed a fine-structure splitting energy of $\Delta = 0.46 \pm 0.03$ meV. For the QD in Figure 3c, we observed the splitting energy between FSS1 and FSS2 of $\Delta_{12} = 0.50 \pm 0.01$ meV and the splitting energy between FSS2 and FSS3 of $\Delta_{23} = 0.84 \pm 0.01$ meV.

To analyze the polarization state of the individual emission peaks, we recorded the PL spectra as a function of $\phi$ (see Figure S2b and S2d) and fitted the evolution of peak intensity (integrated peak area) with respect to $\phi$ using multi-Gaussian functions with the FWHM of each peak as shared free parameter for each individual QD. For example, for the two QDs displayed in Figure 3a and c, the fitted linewidths (FWHM) were $0.21 \pm 0.01$ meV and $0.22 \pm 0.01$ meV, respectively (Figure 3b, d). From the traces, it is visible that all the exciton FSSs of the two single QDs exhibit a noticeable degree of circular polarization (component of π-periodicity), especially in comparison to the purely linear-polarized laser light as displayed in Figure 2b. Similar results were observed for other individual QDs as well (Figure S3, S4).

To quantify the four Stokes parameters: *I, M, C, S*, we fitted the peak intensity trace with Equation (1), where we left the Stokes parameters as free parameters and constrained the polarizer angle, α, and the retardation-plate phase, ξ, within a small range around the experimental values. For a reliable fit result, we carried out the global fit of emission peaks of each individual QD with shared value of α and ξ among the FSSs. Figure 4a shows *S/I* values from all the FSSs of the two QDs displayed in Figure 3, with respect to the energy difference of the fine structure peaks from FSS1. In principle, the detected light has a right-hand circularly polarized (RHCP) component if $\frac{S}{I} > 0$ and a left-hand circularly polarized (LHCP) component if $\frac{S}{I} < 0$. These two QDs show a degree of circular polarization ($\frac{|S|}{I}$) around 0.1-0.4. Over a wide range of QD emission energy (indicating QD of different sizes), $\frac{S}{I}$ values remain rather constant, as shown in Figure 4b. Therefore, we conclude that all measured fine-structure emission peaks possess a non-negligible amount of circularly polarized component with an average value of $\frac{|S|}{I} = 0.16 \pm 0.03$. Interestingly, $\frac{S}{I}$ values change sign among FSS, indicating that the handedness of circular component is flipped with respect to the adjacent energy state. The same observations apply to all the studied QDs as shown in Figure 4b and Figure 4c (upper panel). The observation that within the same QD, some of the FSSs emit σ$^+$(LHCP) light and others σ$^-$(RHCP) light typically appears when external magnetic fields are present,[18,27] which however was not the case in our experiments.

Furthermore, we calculated the degree of linear polarization (DOLP) and the degree of total polarization (DOP) and plotted in the middle and lower panels of Figure 4c, respectively. FSS1,

FSS2 and FSS3 are represented by red, green and blue bars, respectively. Overall, we observed that the DOCP of single QD emission can reach up to ~38%. As for the DOLP, it mostly ranged from 60% to at most 100%, consistent with many experimental studies based on standard polarization measurements which reported a high degree of linear polarization, however not always reaching 100%.[26,30,56] Lower values of DOLP could be attributed to the unresolved peaks for doublets, low intensity-to-noise for the weak peaks within a triplet and fluctuating emission intensity. The total degree of polarization (DOP) often reached near-unity values, suggesting that the DOCP and DOLP fully describes the polarization properties of the exciton FSSs, without unpolarized components.

In conclusion, our results prove that even in the absence of external magnetic fields, the single QD emission possesses a considerable portion of circularly polarized component, in addition to the dominant linearly polarized component, that was already proposed by Isarov *et al.*[57] in magneto-optical studies. We could exclude potential chromatic retardation effects introduced by our experimental setup, since we observed a clear sign change of $\frac{S}{I}$ within an energy window of only a few millielectronvolt within the FSS manifold, after a complete calibration of our setup.

**Discussion**

There are several possibilities for the observed circularly polarized PL in a QD. Circular dichroism and circularly polarized luminescence in QD could for example arise from the introduction of chiral ligand molecules that may impose their enantiomeric structure on the surface of the QD,[58–61] which was also shown for perovskite QDs.[62] Also, bulk and two-dimensional perovskite materials are known to exhibit circular dichroism if activated by chiral molecules.[63–66] Furthermore, circular dichroism was observed in intrinsically chiral QDs, possibly due to the screw dislocations.[10] We can safely exclude both the above-mentioned effects for the circularly polarized luminescence observed in this work, since the QDs in our sample were stabilized with non-chiral zwitterionic ligands. Moreover, such effects do not explain the observed opposite handedness of the FSS emission lines. Neugebauer *et al.*[67] showed that a linear dipole can possess partially circularly polarized component in the non-propagating near-field part of *k*-space. According to this study, if the evanescent longitudinal spin component[68] was coupled from the near-field to a propagating wave by an optically denser medium, it should be detected as circularly polarized light in the far-field.[67] However, the QDs in our measurement were embedded in a polystyrene layer, erasing an abrupt change of the refractive index at QD's vicinity, which excludes this effect. Another possibility is that the coherent coupling of two resonant non-parallel dipoles would also lead to the emission of circularly polarized light. Since the FSSs are non-degenerate and (potentially) thermally populated (no coherent state), coherent coupling among the FSSs can also be ruled out. A similar effect with coherent coupling was also predicted to arise in QD molecules, where two resonant QDs were coupled.[69] However, the simultaneous spectral diffusion of FSS peaks (see Figure S1a) proves that we were detecting single QDs and not two separate, resonant QDs, since the two separate QDs would exhibit independent spectral diffusion trajectories.

Meanwhile, a recent work[70] stated that circular dichroism should also be observable in non-chiral metal halide perovskites, due to a combination of an in-plane symmetry breaking, Rashba splitting, and the effect of the exciton momentum. As discussed earlier, the investigated QDs possess an edge length of ~14 nm and therefore lie in the so-called weak confinement regime. Hence, we do not expect the exciton to have a constant momentum that would induce circular dichroism or circularly polarized emission. On the other hand, the Rashba field itself could be responsible for the circularly polarized component of the fine-structure split states, as recently reported for layered perovskite compounds.[57,71,72] The Rashba effect occurs as a result of an inversion-symmetry breaking in combination with strong spin-orbit coupling.[73] In perovskite QDs this effect was originally considered to be responsible for the large zero-field fine-structure splitting observed in cryogenic PL experiments.[26,37] The $[PbBr_6]^{4-}$ structure is mainly responsible for the electronic band structure and therefore is also responsible for the strong spin-orbit coupling, whereas a displacement of the inorganic $Cs^+$-ion can induce an inversion asymmetry.[74] Charge carriers in perovskite materials should therefore experience a momentum-dependent effective magnetic field,[75] which could result in mixing of bright and dark states[76] or a reduction of the overall symmetry,[18] even in absence of external magnetic field. This may explain why the detected PL of individual FSS was elliptically polarized and the observed change of polarization handedness within the FSS manifold with the lowest energy state emitting $\sigma^-$-polarized light, similar to those results when the external magnetic field was applied.[18] An additional factor could be a reduction of the point symmetry at the level of individual QDs. $D_{2h}$ (the point group related to *Pnma*) has $C_{2v}$ as a potential point subgroup (4 symmetry operations instead of 8), which can induce circular dichroism. A quantitative description requires a more refined theoretical model, where the presented experimental results should serve as motivation and stimulus for further developments.

**Conclusion**

In conclusion, we investigated the polarization properties of the bright exciton FSS at cryogenic temperature by experimentally determining their Stokes parameters. In addition to the dominant linearly polarized component, the emitted light consistently exhibited a non-negligible fraction of circular polarization. Moreover, we observed both LHCP and RHCP light for different FSSs within a given QD. Our results provide an important puzzle piece to gain a complete understanding of the exciton fine-structure of perovskite QDs. Beyond the zero-field regime investigated here, future magnetic-field-dependent measurements of the Stokes parameters could provide additional insight into the evolution of the bright triplet exciton states at higher fields. Furthermore, this elaborate Stokes polarimetric measurement technique can be applied to other types of (quantum) emitters[77,78] or chiral nanostructures[79,80] to close gaps and consolidate the understanding of their photophysical properties with potential application in chiral quantum optics.

ASSOCIATED CONTENT

**Supporting Information**.

In the Supporting Information, we present the PL time-series and the PL spectra of the QDs in the main text as a function of the retardation plate angle. Furthermore, we show PL spectra and their integrated emission as a function of the retardation plate angle for five more QDs (QD#1, QD#3, QD#5, QD#8, QD#10 in Figure 4).

NOTES:

The authors declare no competing financial interest.

AUTHOR INFORMATION

ORCID:

| | |
|---|---|
| **Virginia Oddi:** | 0009-0005-3962-3664 |
| **Chenglian Zhu:** | 0000-0001-8638-6925 |
| **Michael A. Becker:** | 0000-0003-2042-9384 |
| **Dmitry N. Dirin:** | 0000-0002-5187-4555 |
| **Taehee Kim:** | 0000-0001-9426-1006 |
| **Rainer F. Mahrt:** | 0000-0002-9772-1490 |
| **Jacky Even:** | 0000-0002-4607-3390 |
| **Gabriele Rainò:** | 0000-0002-2395-4937 |
| **Makysm V. Kovalenko:** | 0000-0002-6396-8938 |
| **Thilo Stöferle:** | 0000-0003-0612-7195 |

ACKNOWLEDGMENT

C.Z., G.R., T.S. acknowledge funding from the Swiss National Science Foundation (Grant No. 200021_192308, "Q-Light – Engineered Quantum Light Sources with Nanocrystal Assemblies"). V.O., T.S. and R.F.M. acknowledge funding from the EU H2020 MSCA-ITN "PERSEPHONe" (Grant No. 956270). This work was also supported by the Air Force Office of Scientific Research and the Office of Naval Research under award number FA8655-21-1-7013.

**FIGURES**

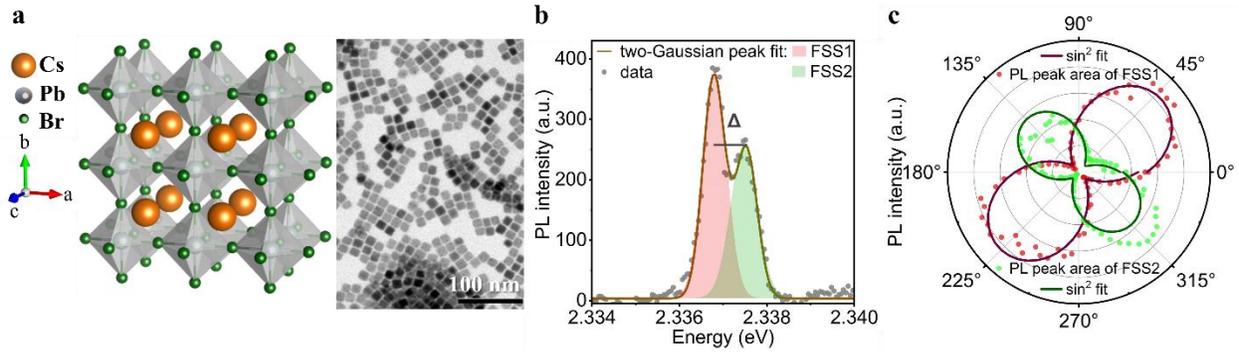

**Figure 1**: **Crystal structure and emission properties of single cesium lead bromide QDs at 4 K. a**, Schematic crystal structure of CsPbBr$_3$ QDs (left) and high-resolution transmission electron micrograph of CsPbBr$_3$ QDs (right). **b**, PL spectrum of a single CsPbBr$_3$ QD at 4 K. The brown solid line is the fit of two-Gaussian-peak function. This QD exhibits a doublet exciton with a splitting energy of Δ=0.75 ± 0.08 meV. **c**, Peak-area-intensity of each Gaussian fit as a function of linear polarizer angle for the QD displayed in **b**. Solid lines represent sin$^2$ fits to the data, revealing a DOLP of 82.9 ± 2.2% and 78.6 ± 4.8% for FSS1 and FSS2, respectively.

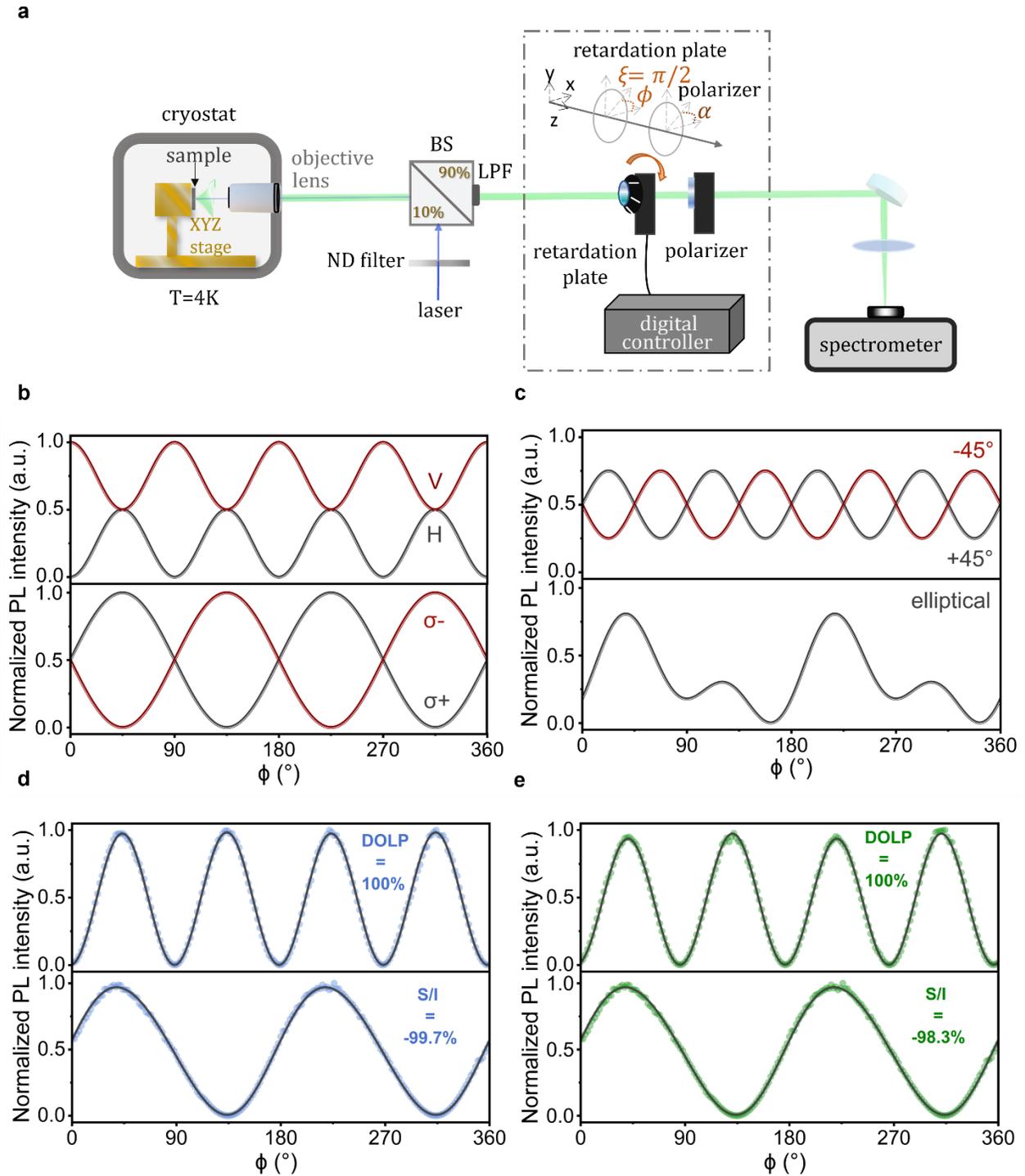

**Figure 2**: **Experimental setup and examples of Stokes polarimetric measurements for various defined polarizations. a**, Schematic of the experimental setup. The inset depicts the basic measurement unit consisting of a *λ/4*-waveplate and a linear polarizer to perform Stokes polarimetric measurements. **B**, Calculated intensity variation for fully linearly polarized light along the horizontal (H) and vertical (V) direction (upper panel) and right-hand (σ⁻) circularly polarized

light (lower panel) as a function of the λ/4-waveplate angle, ϕ. **c**, Calculated intensity variation for fully linearly polarized light along the +45°/-45° direction (upper panel) and elliptically polarized light (lower panel) as a function of the λ/4-waveplate angle, ϕ. **d**, Detected intensity variation as a function of λ/4-waveplate angle, ϕ. The defined linear (upper panel) and circular (lower panel) polarization are obtained from the laser light, generated with a combination of linear polarizer and λ/4-waveplate. Solid grey lines represent a fit of Equation (1) to the data from which we retrieve a value of DOLP equal to 100% and a $\frac{S}{I}$ of -99.7% for linear and circular polarization, respectively. **E,** Detected intensity variation as a function of λ/4-waveplate angle, ϕ, for the linearly (upper panel) and circularly (lower panel) polarized laser light in **d** after going through our experimental setup. From the fit of Equation (1) to the data, we obtain a DOLP of 100% and a value of $\frac{S}{I}$ equal to -98.3%. Comparing **d** and **e**, we prove that our experimental setup produces no or only negligible artifacts on the detected results.

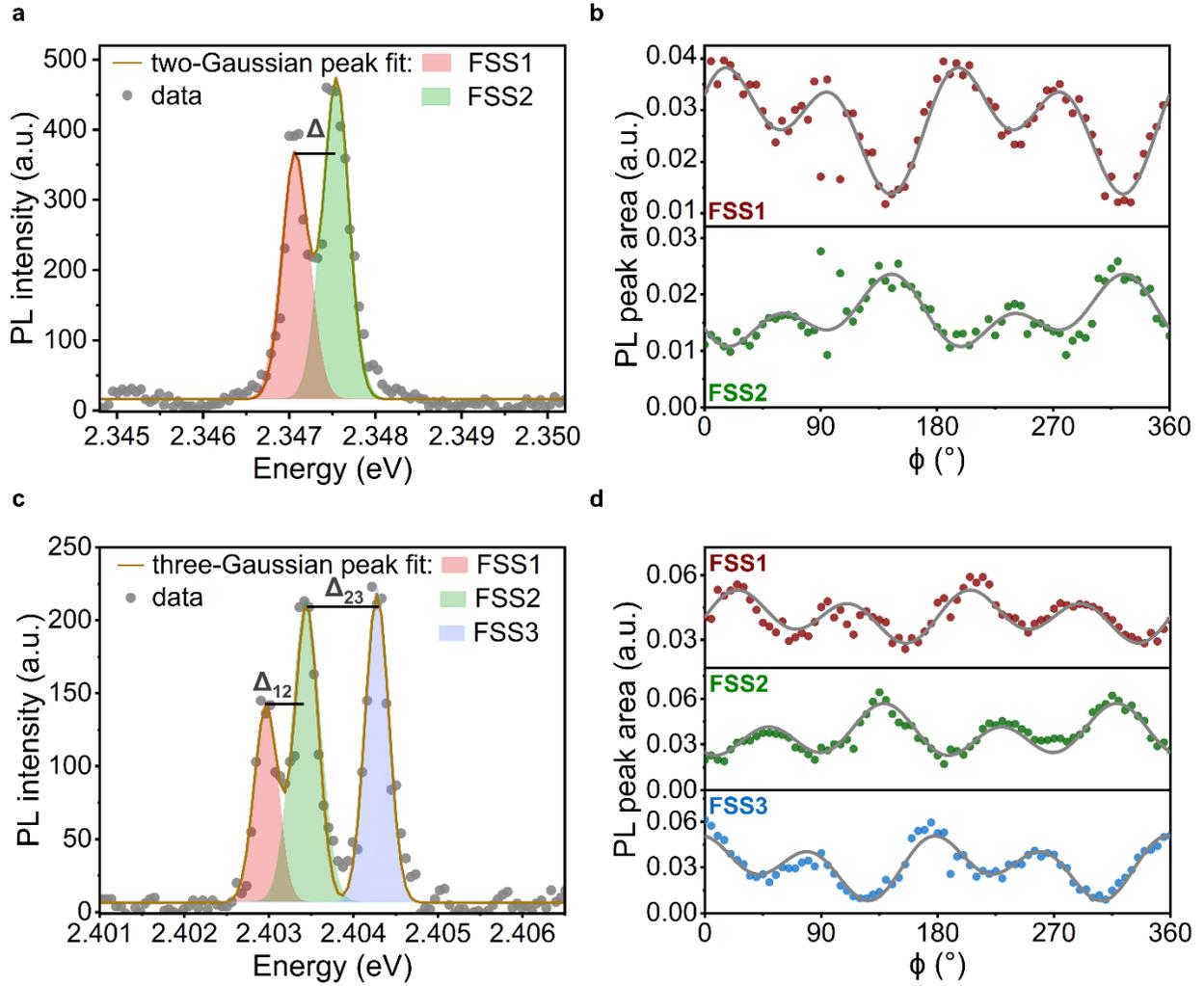

**Figure 3**: **Stokes polarimetric measurements on exciton FSS in single CsPbBr$_3$ QDs at 4 K. a**, Photoluminescence spectrum of a single CsPbBr$_3$ QD exhibiting a doublet exciton FSS with a splitting Δ. The brown solid line is a two-Gaussian peak fit to the data. **b**, Peak-area-intensities of the doublet exciton fine-structure from Gaussian fits of the QD shown in **a,** as a function of λ/4-waveplate angle, ϕ. Solid grey lines represent a fit of Equation (1) to the data. **c**, PL spectrum of a single CsPbBr$_3$ QD exhibiting a triplet exciton fine-structure with splitting $\Delta_{12}$ and $\Delta_{23}$ between FSS1 and FSS2 and between FSS2 and FSS3, respectively. The brown solid line is a three-Gaussian peak fit to the data. **d**, Peak-area-intensities of the triplet exciton fine-structure from Gaussian fits of the QD shown in **c,** as a function of λ/4-waveplate angle, ϕ, for FSS1 (upper panel), FSS2 (middle panel) and FSS3 (lower panel). Solid grey lines represent a fit of Equation (1) to the data.

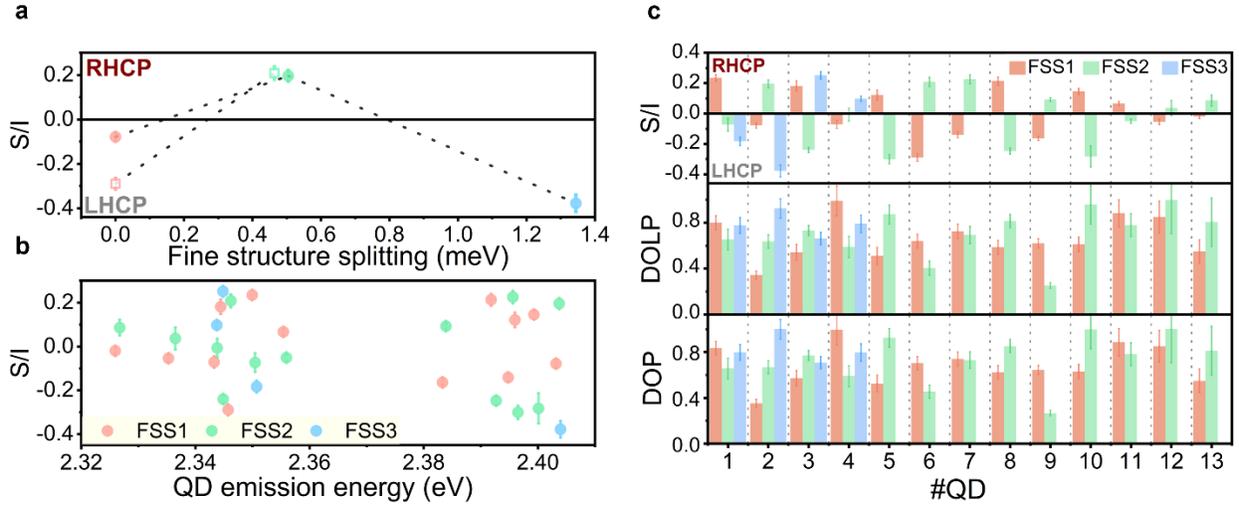

**Figure 4**: **Measured Stokes parameters from each of the exciton fine-structure states in different single CsPbBr$_3$ QDs**. **a**, Exemplary $\frac{S}{I}$ values for two QDs exhibiting doublet (open squares) and triplet (filled spheres) exciton as a function of fine-structure splitting relative to FSS1, displayed in Figure 3a-b and Figure 3c-d, respectively. **b**, $\frac{S}{I}$ of all FSSs versus emission energy of individual QDs. **c**, Extracted $\frac{S}{I}$ values, degree of linear polarization (DOLP) and total degree of polarization (DOP) for all the exciton fine-structure peaks in different QDs. The red bars correspond to FSS1, and green bars correspond to FSS2. When the complete fine-structure triplet was resolved, the blue bar represents FSS3. QD number 2 and 6 are the QDs discussed in the main text. The error bars are obtained from the standard errors of the fit parameters when fitting Equation (1) to the measured data.

# Supporting Information

# Circularly Polarized Luminescence Without External Magnetic Fields from Individual CsPbBr$_3$ Perovskite Quantum Dots

*Virginia Oddi[1,2,#], Chenglian Zhu[2,3,#], Michael A. Becker[1,$], Yesim Sahin[2,3], Dmitry N. Dirin[2,3], Taehee Kim[2,3], Rainer F. Mahrt[1], Jacky Even[4], Gabriele Rainò[2,3,*], Maksym V. Kovalenko[2,3,*], Thilo Stöferle[1,*]*

[1] IBM Research Europe – Zurich, Säumerstrasse 4, 8803 Rüschlikon, Switzerland.

[2] Institute of Inorganic Chemistry, Department of Chemistry and Applied Biosciences, ETH Zurich, 8093 Zurich, Switzerland.

[3] Laboratory for Thin Films and Photovoltaics, Empa, Swiss Federal Laboratories for Materials Science and Technology, 8600 Dübendorf, Switzerland.

[4] Université de Rennes, INSA Rennes, CNRS, Institut FOTON - UMR6082, 35000 Rennes, France.

**\* Corresponding Authors:** rainog@ethz.ch**,** mvkovalenko@ethz.ch, tof@zurich.ibm.com**,**
[#] These authors equally contributed to this work.
[$] Present address: Zeiss SMT, Oberkochen, Germany

## Contents

# I. Materials and Methods

## 1.1 Chemicals

Cesium carbonate ($Cs_2CO_3$, 99.9%), 1-octadecene (ODE, 90%), 3-(N,N-dimethyloctadecylammonio) propanesulfonate (ASC18, >99.0%) and oleic acid (90%) were purchased from Sigma-Aldrich; lead(II) acetate trihydrate (>99%, for analysis) and bromine ($Br_2$, >99%) from Acros Organics; trioctylphosphine (TOP) from Strem; toluene (99.85%, extra dry over molecular sieve, AcroSeal®) from Thermo Scientific; and ethyl acetate (EtOAc, >99.7%, HPLC grade) from Fisher Scientific.

## 1.2 Precursor Syntheses

**Cesium oleate:** $Cs_2CO_3$ (1.628 g, 5 mmol) and oleic acid (5 mL, 16 mmol) were evacuated in a three-neck flask along with 20 mL of ODE at room temperature until the first gas evolution stops, heated to 120 °C under vacuum, and then further evacuated for 1 hour at this temperature. This yields a 0.4 M solution of Cs-oleate in ODE. The solution turns solid when cooled to room temperature and was stored under nitrogen and heated before use.

**Lead-oleate:** Lead (II) acetate trihydrate (4.6066 g, 12 mmol) and oleic acid (7.6 mL, 24 mmol) were evacuated in a three-neck flask along with 16.4 mL of ODE at room temperature until the first gas evolution stops, heated to 120 °C under vacuum, and then further evacuated for 1 hour. This yields a 0.5 M solution of Pb-oleate in ODE. The solution turns solid when cooled to room temperature and was stored under nitrogen and heated before use.

**TOP-$Br_2$:** In a Schlenk flask, TOP (6 mL, 13 mmol) was dissolved in 18.7 ml anhydrous toluene and $Br_2$ (0.6 mL, 11.5 mmol) was added dropwise with vigorous stirring. The resulting solution was stirred for 1 hour under a nitrogen atmosphere, forming a white-pale yellow viscous solution at the end. The solution (app. 0.46 M) was stored under nitrogen and heated before use.

## 1.3 $CsPbBr_3$ QD Synthesis

107.5 mg ASC18, 2.0 ml Cs-oleate, 2.5 ml Pb-oleate and 5 ml ODE was added to a 100 ml three-neck round-bottom flask with a stir bar. The reaction mixture was evacuated and refilled with nitrogen 3 times, then was heated to 180°C under nitrogen atmosphere followed by the injection of 2.5 ml TOP-$Br_2$ with vigorous stirring. The resulting solution was rapidly cooled to room temperature using an ice-water bath and subjected to centrifugation at 12.1 krpm for 10 minutes. The initial precipitate obtained from this crude solution was size fractioned in multiple cycles by adding various amounts of toluene for dispersion (table below). In each cycle, dispersion was centrifuged at 12.1 krpm for 10 minutes, followed by collecting the supernatant and redispersing the precipitate in toluene again. The supernatants from these cycles were denoted as $pf_n$ (where n=1, 2, 3, etc.). The supernatant from the 6[th] cycle ($pf_6$) was washed once with 2 equivalents of EtOAc. The resulting precipitate was transferred directly into the glovebox and redispersed in 1 ml of anhydrous toluene. The final solution was filtered using 0.45 μm syringe filter and used in the experiments.

| Fraction | Pf1 | Pf2 | Pf3 | Pf4 | Pf5 | Pf6 |
|---|---|---|---|---|---|---|
| Amount of Toluene Used for Dispersion | 1 ml | 1 ml | 1 ml | 2 ml | 2 ml | 2 ml |

## 2. Sample preparation

The single QD samples were prepared in a glovebox that is kept under a nitrogen atmosphere. The colloidal dispersion with a concentration of ~1 mg/ml was diluted by a factor 100 in toluene (Acros Organics, 99.85% extra dry over molecular sieve). The solution is further diluted by another factor 100 in a 3-mass% solution of polystyrene (ALDRICH, average Mw ~280,000) in toluene, whereupon the solution was spin-coated at 3000 rpm. for 60 s onto a crystalline Si wafer covered with a 3-µm-thick thermal-oxide layer.

### 3.1 Characterization of CsPbBr$_3$ solution

**Absorption spectra (UV-Vis):** Optical characterizations were performed at ambient conditions. UV-Vis absorption spectra of colloidal NCs were collected using a Jasco V670 spectrometer in transmission mode.

**Photoluminescence (PL):** A Fluorolog iHR 320 Horiba Jobin Yvon spectrofluorometer equipped with a PMT detector was used to acquire steady-state PL spectra. NC solutions were measured in the same dilutions and solvents as the absorption measurements.

**Transmission electron microscopy (TEM):** The images were recorded using a JEOL JEM-1400+ microscope operated at 120 kV. Images were processed using ImageJ.

### 3.2 Single QD optical characterization

For single-QD spectroscopy, a home-built µ-PL setup was used. Samples were mounted on a xyz nano-positioning stage inside an evacuated liquid-helium closed-loop cryostat (MONTANA INSTRUMENTS) and cooled down to a targeted temperature of 4 K. Single QDs were excited by means of a fiber-coupled excitation laser, which was focused (Gaussian spot with $1/e^2$ diameter of 2.4 µm) on the sample by a dry microscope objective (NA = 0.8, 100×). Typical fluences used to excite single QDs were in the range 2-6 nJ/cm$^2$. The emitted light was collected by the same objective and passed through a 90:10 beam splitter and a long-pass filter at 500 nm. A monochromator coupled to an EMCCD (Princeton Instruments, 0.75-m) was used for recording the spectra. PL spectra were measured with a grating of 1800 lines/mm, blaze at 500 nm (0.2 meV spectral resolution). For the Stokes polarimetric measurement, we used a combination of linear polarizer and retardation plate (λ/4-wave) in the detection path.

## 4. Calibration of the setup

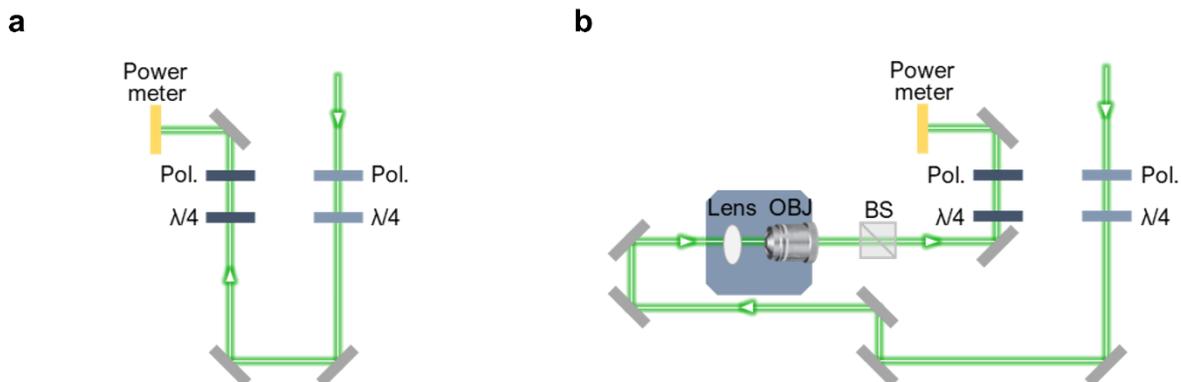

**Figure S1: a,** A well-defined linear and circular polarization of the laser light was prepared by using either a linear polarizer or a combination of linear polarizer and λ/4-waveplate. The laser polarization state was analyzed through a rotating λ/4-waveplate and a fixed linear polarizer.. Transmitted intensity as a function of the λ/4-waveplate angle, $\phi$, was recorded by a power meter. **b,** We mimicked the PL optical path to assess the extent of modulation introduced by the optical components in the PL detection path. The polarized laser light set in **a** was sent through the cryostat, and then through all the optical components in the detection path of our setup (microscope objective lens, BS, LPF, mirrors, and lenses). Then, its polarization state was analyzed through a rotating λ/4-waveplate and a fixed polarizer. Again, the transmitted light intensity as a function of the λ/4-waveplate angle, $\phi$, was recorded by a power meter.

## II. 2D plot of time-series and Stokes measurements

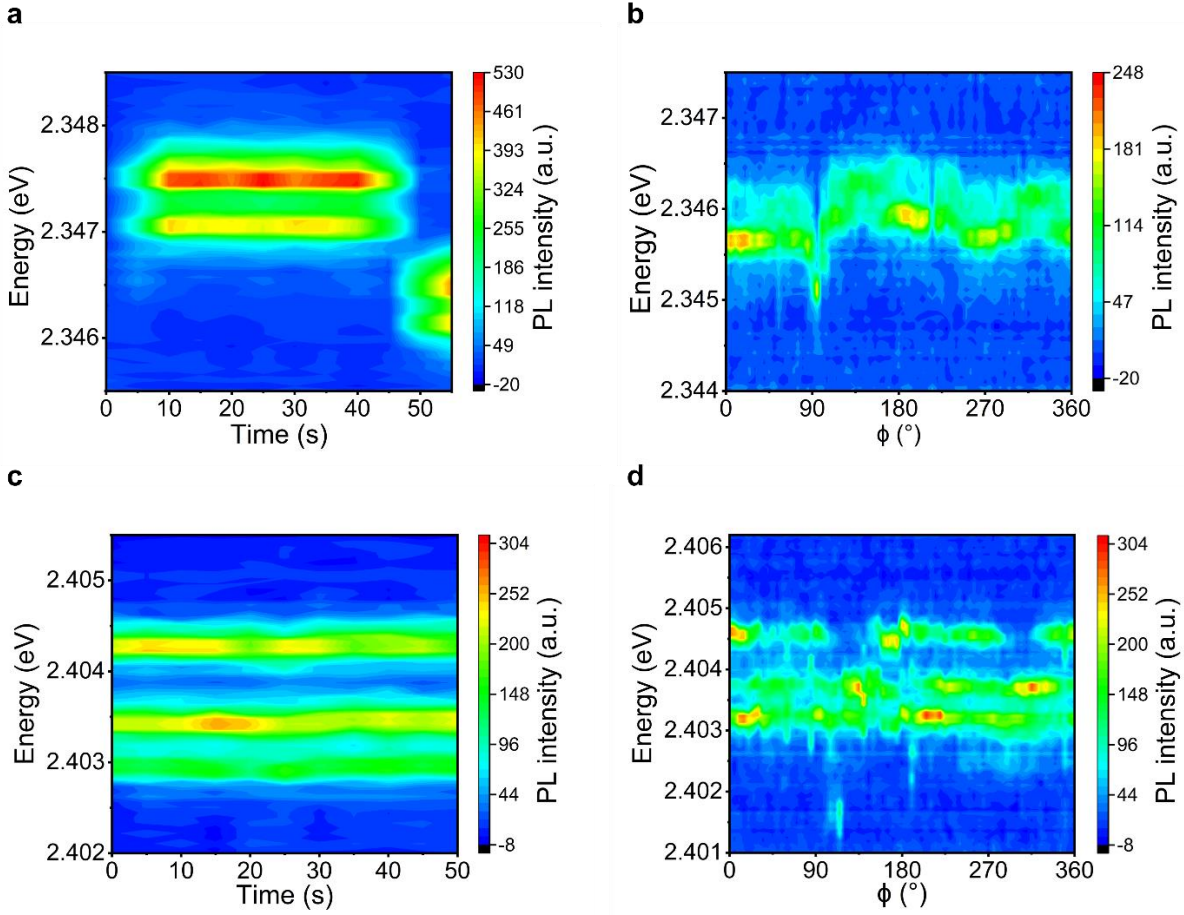

**Figure S2: Time-series and spectral diffusion: a,** PL time-series of the QD displayed in Figure 3a, acquired with an integration time of 5 s. **b,** PL spectra as a function of the λ/4-waveplate angle, $\phi$, for the QD displayed in Figure 3a, acquired with an integration time of 5 s. **c,** PL time-series of the QD in Figure 3c, acquired with an integration time of 5 s. **d**, PL spectra as a function of the λ/4-waveplate angle, $\phi$, of the QD in Figure 3c, acquired with an integration time of 10 s.

# III. Fitting of Stokes parameters

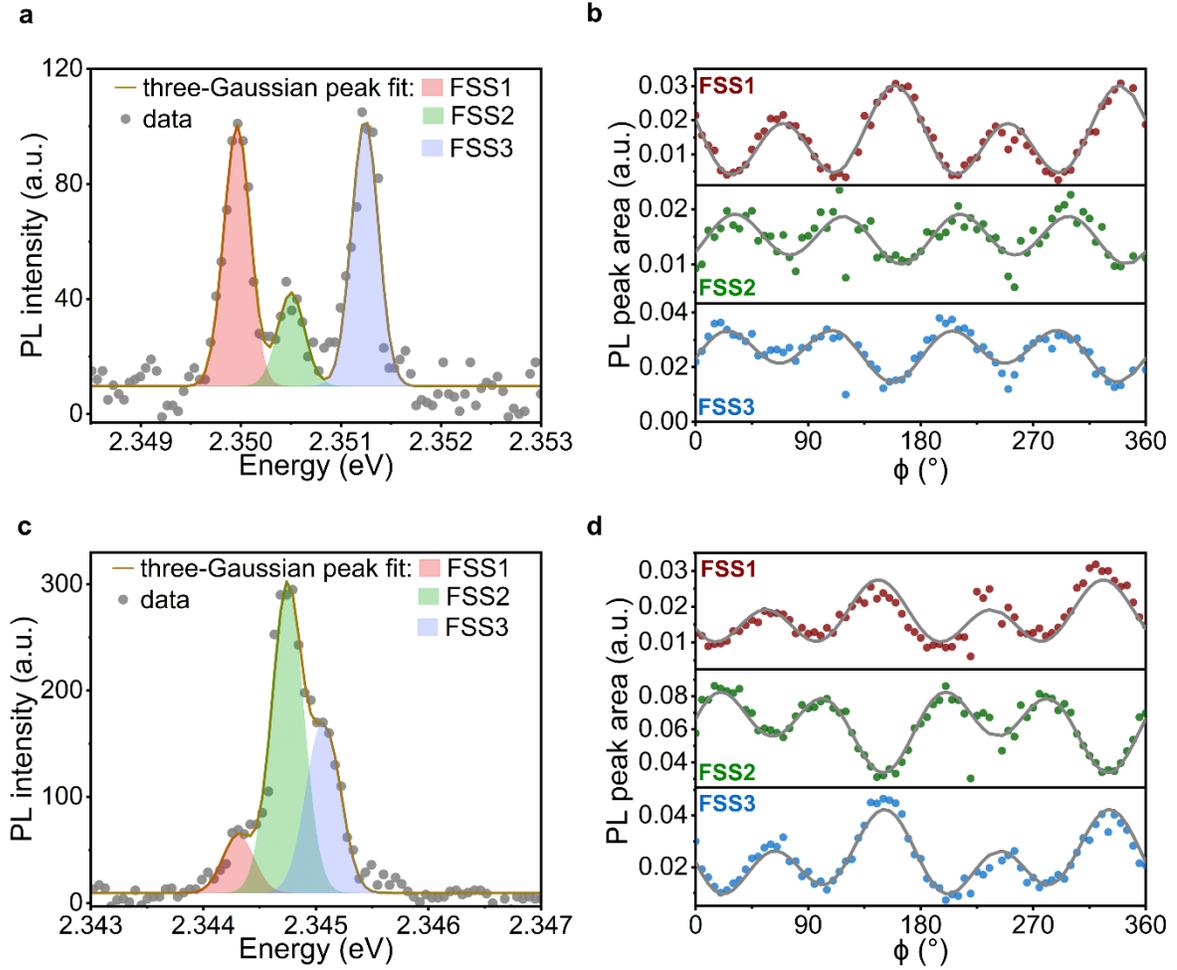

**Figure S3: Stokes polarimetric measurements on QD#1 and QD#3 with triplet exciton. a**, Spectrum of QD#1 in Figure 4. Triplet exciton is fitted with the sum of three Gaussian functions. **b**, Peak-area-intensity of the individual FSS of QD#1, extracted from the Gaussian peak fits, as a function of the λ/4-waveplate angle, ϕ. **c,** Spectrum of QD#3 in Figure 4. Triplet exciton is fitted with the sum of three Gaussian functions. **d**, Peak-area-intensity of the individual FSS in QD#3, extracted from the Gaussian peak fits, as a function of the λ/4-waveplate angle, ϕ.

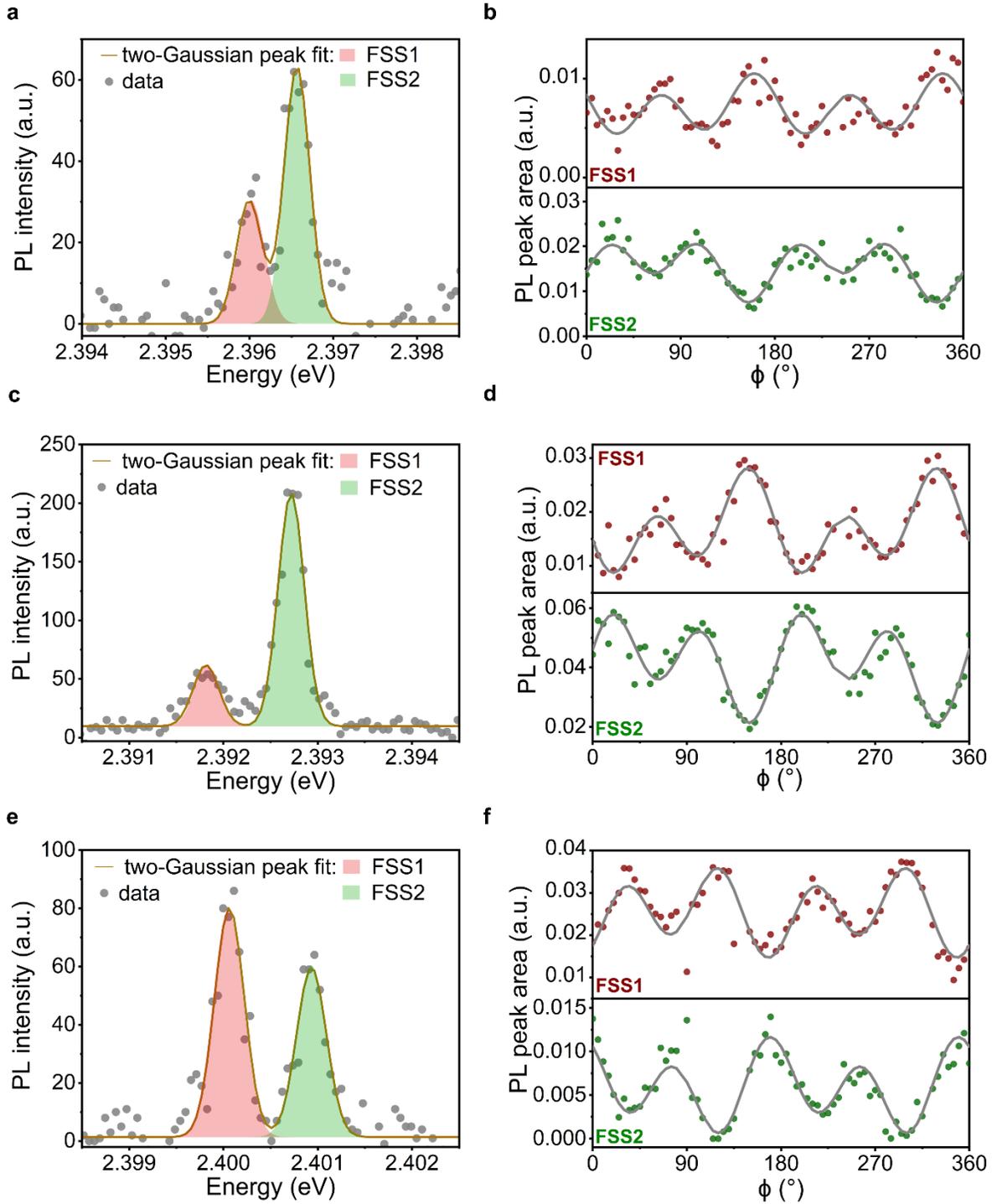

**Figure S4: Stokes polarimetric measurements on QD#5, QD8 and QD#10 with doublet exciton. a**, Spectrum of QD#5 in Figure 4. Doublet exciton is fitted with the sum of two Gaussian peaks. **b**, Peak-area-intensity of the individual fine-structure peaks in QD#5, extracted from the Gaussian peak fits as a function of the λ/4-waveplate angle, ϕ. **c** Spectrum of QD#8 in Figure 4. Doublet exciton is fitted with the sum of two Gaussian peaks. **d**, Peak-area-intensity of the individual fine-structure peaks in QD#8, extracted from the Gaussian peak fits as a function of the λ/4-waveplate angle, ϕ. **e**, Spectrum of QD#10 in Figure 4. Doublet exciton is fitted with the sum

of two Gaussian peaks. **f**, Peak-area-intensity of the individual fine-structure peaks in QD#10, extracted from the Gaussian peak fits as a function of the λ/4 angle, ϕ.